\documentclass[prb,aps,twocolumn,floatfix]{revtex4}
\usepackage{graphicx,bm}
\usepackage{subfigure}
\usepackage{comment}
\usepackage{amssymb}
\usepackage{verbatim}

\newcommand{\avg}[1]{\langle #1 \rangle}

\newcommand{\nn}{\nonumber\\}
\newcommand{\up}{\uparrow}
\newcommand{\down}{\downarrow}

\newcommand{\midb}[1]{\left[ #1 \right]}

\newcommand{\half}{{1\over 2}}
\begin{document}
\pagestyle{plain}
\title{Macrospin Models of Spin Transfer Dynamics}
\author{Jiang Xiao and A. Zangwill}
\affiliation{School of Physics, Georgia Institute of
Technology, Atlanta, GA 30332-0430}
\author{M. D. Stiles}
\affiliation{Electron Physics Group, National Institute of Standards and
Technology, Gaithersburg, MD 20899-8412}
\begin{abstract}
    The current-induced magnetization dynamics of a spin valve are studied using a macrospin (single domain) approximation and numerical solutions of a generalized Landau-Lifshitz-Gilbert equation. For the purpose of quantitative comparison with experiment [Kiselev {\it et al.} Nature {\bf 425}, 380 (2003)], we calculate the resistance and microwave power as a function of current and external field including the effects of anisotropies, damping, spin-transfer torque,  thermal fluctuations, spin-pumping, and incomplete absorption of transverse spin current. While many features of experiment appear in the simulations, there are two significant discrepancies: the current dependence of the precession frequency and the presence/absence of a microwave quiet magnetic phase with a distinct magnetoresistance signature. Comparison is made with micromagnetic simulations designed to model the same experiment.

\end{abstract}
\date{\today}
\maketitle
\section{Introduction}

\begin{figure}
    \centering
    \includegraphics[width= 2.75in]{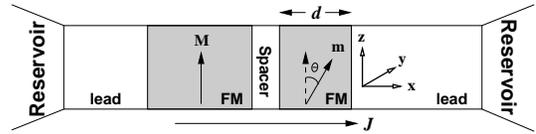}
    \caption{Side view of a spin-valve (schematic). A non-magnetic spacer layer is sandwiched between a ``fixed'' ferromagnetic film with uniform magnetization ${\bf M}$ and a ``free'' ferromagnetic film with uniform magnetization ${\bf m}$. The leads on either side of the sandwich are non-magnetic. The free layer has thickness $d$ and the direction of positive electric (negative electron) current $J>0$ is indicated.}
    \label{fig:spinvalve}
\end{figure}

In the ten years since Slonczewski\cite{Slonczewski:1996} and Berger\cite{Berger:1996} discovered the quantum mechanical phenomenon of spin-transfer torque,\cite{StilesMiltat} considerable evidence has accumulated that a spin-polarized current that passes through a thin ferromagnetic film can induce switching and/or precession of the film's magnetization. Early experiments used multilayers,\cite{Tsoi:1998} nanowires,\cite{Wegrowe:1999} small particle junctions,\cite{Sun:1999} and point contacts\cite{Myers:1999} to infer the presence of this effect. The most convincing data came later from pillar-type ``spin valves'' with nanometer-scale transverse dimensions (Fig.\ref{fig:spinvalve}) where a thin film non-magnet is sandwiched between two thin film ferromagnets.\cite{Katine:2000b,Grollier:2001} In the range of film thicknesses most commonly used, the magnetization ${\bf M}$ of the thick ``fixed'' layer and the magnetization ${\bf m}$ of the thin ``free'' layer lie in the plane of the film. Non-magnetic leads connect the spin valve to electron reservoirs.

Due to the phenomenon of giant magnetoresistance,\cite{Bass:1999} voltage measurements are sufficient to reveal that hysteretic switching of ${\bf m}$ occurs as a function of the applied current density $J$ when a magnetic field $H$ smaller than the coercive field is applied along the easy axis of the free layer.
For larger values of $H$, it is believed that ${\bf m}$ exhibits one or more types of stable precession as a function of $J$ until the current density is large enough to induce switching. This conclusion\cite{Kiselev:2003,Rippard:2004a,Covington:2004,Rippard:2004b,Krivorotov:2005} is based on the  experimental observation of  narrow band microwave emission combined with  calculations using a generalized Landau-Lifshitz-Gilbert (LLG) equation
that predict precession of the free layer.
Other observed dynamical behavior includes telegraph noise that is interpreted as rapid switching between two distinct states of magnetization.\cite{Myers:2002,Urazhdin:2003,Pufall:2004,Krivorotov:2004}

Several experimental groups have used the macrospin (single domain) approximation to propose ``phase diagrams'' that identify the dynamical state of their spin valves as a function of $J$ and $H$.\cite{Katine:2000b,Sun:2003,Grollier:2003,Kiselev:2003,Urazhdin:2003,Krivorotov:2004,Zimmler:2004,Lacour:2004}  There have also been purely theoretical studies of the  LLG equation (generalized to include spin-transfer torque) using both macrospin models\cite{Sun:2000,Bazaliy:1998,Bazaliy:2001,Bazaliy:2004,Xi:2004,Morise:2005,Russek:2005} and micromagnetics simulations\cite{Miltat:2001,Li:2003,Zhu:2004,Lee:2004,Montigny:2005,Berkov:2005,Berkov:2005b,Xi:2005} that do not make the single-domain approximation. Unfortunately, it is difficult to extract a coherent picture from all this work because different authors make different choices for the physical effects they believe most affect the dynamics. There is not even unanimity amongst authors for the form of the spin-transfer torque itself.

This state of affairs motivated us to perform a thorough study of the LLG dynamics of a model spin-valve for the purpose of a quantitative comparison with the data reported by Kiselev {\it et al.}\cite{Kiselev:2003} for a Co/Cu/Co nanopillar. We make the macrospin approximation, but otherwise systematically examine the effects of  different forms of spin-transfer torque,  thermal fluctuations, spin-pumping, incomplete absorption of transverse spin current, and angle-dependent damping. We find that a ``minimal'' macrospin model can reproduce many (but not all) features of the experiment. The most important points of disagreement are the current dependence of the precession frequency and the existence of a microwave quiet magnetic phase with a distinct magnetoresistance signature. In light of these results, we comment on micromagnetic simulations\cite{Lee:2004,Berkov:2005b} designed to model the identical set of experimental data.

The plan of this paper is as follows. Section II describes the macrospin models of interest and the generalized Landau-Lifshitz-Gilbert equation we solve numerically. Section III presents results for a ``minimal'' model and compares them to the measurements reported in Ref.~\onlinecite{Kiselev:2003}. Section IV examines several variations of the minimal model within the context of the macrospin approximation. Section V compares our results with micromagnetic simulations. Section VI summarizes our results vis \`{a} vis experiment. An Appendix provides some  details omitted from the main body of the paper.

\section{The Macrospin Model}
\begin{figure}
    \centering
    \includegraphics[scale=0.65]{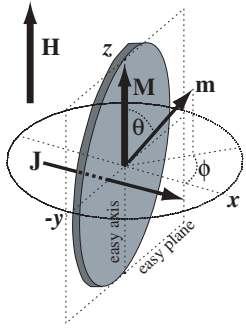}
    \caption{The ellipsoidal cross section of the free layer (shaded) lies in $y$-$z$ plane. We represent its  magnetization by a macrospin ${\bf m}$ that can point in any direction. The fixed layer (not shown) is represented by a fixed macrospin ${\bf M}\parallel {\bf \hat{z}}$.}
    \label{fig:geometry}
\end{figure}

Our macrospin model of the spin valve shown in Fig.~\ref{fig:spinvalve} assumes that the
magnetization is spatially uniform in both ferromagnetic layers with saturation value $M_{\rm s}$. The fixed layer magnetization is ${\bf M}=M_s\hat{\bf z}$, but we allow the unit vector in the direction of the free layer magnetization ${\bf \hat m}={\bf m}/M_s$  to point in any direction. In the coordinate system used here (Fig.~\ref{fig:geometry}),
\begin{equation}
\label{mM}
    {\bf \hat{m}} = \hat{\bf x}\sin\theta\cos\phi
              + \hat{\bf y}\sin\theta\sin\phi +\hat{\bf z}\cos\theta.
\end{equation}
The experiments of interest\cite{Kiselev:2003} use a ``free'' ferromagnetic layer with a thickness $d\approx 3$ nm and an elliptical shape of dimensions about $130$ nm $\times$ $70$ nm. Under these conditions, magnetostatic shape anisotropy makes the $y$-$z$ plane an easy plane for ${\bf m}$. The $z$-axis is an easy axis in that plane. The control parameters are an external magnetic field $H$ directed along $+z$ and an electric current $J$ that is reckoned  positive when negatively charged electrons flow from $+x$ to $-x$.

We describe the dynamics of ${\bf \hat m}$ using a generalized Landau-Lifshitz-Gilbert (LLG) equation,\cite{Slonczewski:1996,Smith:2004}
\begin{equation}
    {d{\bf \hat m}\over dt}
    =  -\gamma {\bf \hat m}\times[ {\bf H}_{\rm eff} + {\bf H}_{\rm T}] + \alpha\hspace{0.05em} {\bf \hat m}\times {d {\bf \hat m}\over dt}+ {\gamma\over \mu_0M_s}{\bf N}.
    \label{eqn:Gilbert}
\end{equation}
It will be convenient to discuss each term in Eq.~(\ref{eqn:Gilbert}) in turn.

\subsection{Energy}
The  first term on the right side of Eq.~(\ref{eqn:Gilbert}) is a conventional magnetic torque with gyromagnetic ratio $\gamma$. This torque is driven by an effective field derived from the total energy $E$ of the free layer with volume $V$:
\begin{equation}
    \mu_0{\bf H}_{\rm eff}=-{1\over V}{\partial E \over \partial {\bf m}}.
\end{equation}
Taking account of magnetostatics, the external field, and a uniaxial surface anisotropy, we show in Appendix I that $E$ can be written in the form
\begin{eqnarray}
    {2 E\over \mu_0 M_{\rm s}^2 V}
       &=& h_Z\cos^2\theta + h_Y\sin^2\theta\sin^2\phi + h_X\sin^2\theta\cos^2\phi \nn
       &-& 2h\cos\theta.
    \label{eqn:E_total2}
\end{eqnarray}
Here, $h=H/M_s$ and the constants $h_X$, $h_Y$, and $h_Z$ are computed in Appendix I using the free layer data given just below Eq.~(\ref{mM}) and the material constants listed in Table I.
\subsection{Damping}
\label{damping}
The ``Gilbert damping'' term $\alpha {\bf \hat{m}}\times \dot{\bf \hat{m}}$ in Eq.~(\ref{eqn:Gilbert}) takes account of energy dissipation mechanisms such as coupling to lattice vibrations\cite{Suhl:1998} and spin-flip scattering.\cite{Kambersky:1970} The prefactor $\alpha$ is usually treated as a phenomenological constant (Table I) although it is not known whether this is a good approximation for situations where the amplitude of precessional motion is large. The Landau-Lifshitz approach to damping replaces the Gilbert term in Eq.~(\ref{eqn:Gilbert}) by
\begin{equation}
      \lambda {\bf \hat m}\times ({{\bf \hat m}\times {\bf H}_{\rm eff}}).
    \label{eqn:LLDamping}
\end{equation}
The constant $\lambda$ can be calculated in some microscopic models, \cite{Fredkin:2000} but a  phenomenological treatment is almost universal. When ${\bf N}=0$ in Eq.~(\ref{eqn:Gilbert}), the Gilbert and Landau-Lifshitz expressions for the damping torque are known to be equivalent, at least formally.\cite{Mallinson:1987} Section~\ref{sts} gives a reason why we prefer the Gilbert form, but we performed calculations using both forms for purposes of comparison. No significant differences were found.

\subsection{Thermal Fluctuations}
The stochastic vector ${\bf H}_{\rm T}$ in Eq.~(\ref{eqn:Gilbert}) is used to simulate the effect of finite temperature. Each Cartesian component is chosen at random from a normal distribution with a variance chosen so the system relaxes to a Boltzmann distribution at equilibrium.\cite{Brown:1963} Specifically,
\begin{equation}
    \avg{H_{\rm T}^i(t)H_{\rm T}^j(t')} =  {2k_B T\alpha\over \gamma V \mu_0M_{\rm s}}\delta_{ij}\delta(t-t'),
\end{equation}
where $i, j = x,y,z$. We have confirmed numerically that this procedure does indeed produce
a Boltzmann distribution of energies at temperature $T$ when ${\bf N}=0$ in Eq.~(\ref{eqn:Gilbert}).  Details of our implementation of the stochastic contributions are indicated in Sec.~\ref{CD}.
\begin{table}
    \centering
    \begin{tabular}{|l|l|}
    \hline
    Quantity            & Values\\
    \hline\hline
    $M_{\rm s}$         & $\rm 0.127\times 10^7~A/m$ [\onlinecite{Kiselev:2003}]\\
  $\mu_0 M_{\rm s}$     & $\rm 1.6~{\rm T}$ \\
    $\rm\gamma(Co)$     & $\rm 2.4 \times 10^5 \,\,m/(A$$\cdot$$\rm s)$  [\onlinecite{Bhagat:1974}]\\
    $\rm \alpha(Co)$    & $0.01$ [\onlinecite{McMichael:2005}] \\
    $K_u$               & $\rm 0.5\times 10^{-3}~J/m^2$ [\onlinecite{Bruno:1989}]\\ \hline
    $ g^{\up\down}/S\, \rm{(Cu)}$   & $\rm 2.94 \times 10^{19}~m^{-2}$ [\onlinecite{Tserkovnyak:2005}]\\
    $\rm \nu \,(Co/Cu)$ & $\rm 0.98$  [\onlinecite{Tserkovnyak:2005}]\\ \hline
    \end{tabular}
    \caption{Quantity Values}
    \label{tab:values}
\end{table}
\subsection{Spin-Transfer}
\label{sts}
The quantity ${\bf N}$ in Eq.~(\ref{eqn:Gilbert}) stands for one of several torque densities that arise from microscopic considerations of the transport of electrons through a spin valve. The most important of these is the spin-transfer torque density ${\bf N}_{\rm st}$.\cite{Slonczewski:1996,Berger:1996,StilesMiltat} A variety of theoretical methods confirm the following picture.\cite{variety} The current that flows through the spin valve shown in Fig~\ref{fig:spinvalve} is spin-polarized. Because ${\bf M}$ and ${\bf m}$ are not collinear, the conduction electron spins that encounter the free layer generally possesses a component of angular momentum that is transverse to the magnetization of free layer itself. Realistic calculations show that this transverse component of angular momentum is largely absorbed by the ferromagnet.\cite{Stiles:2002b,Xia:2002}  Since we describe the free layer as a uniformly  magnetized particle, the absorbed angular momentum generates a torque that appears on the right hand side of  Eq.~(\ref{eqn:Gilbert}). According to current theory,\cite{Slonczewski:2002,Kovalev:2002,Xiao:2004,Fert:2004,Manschot:2004}
\begin{equation}
    {\bf N}_{\rm st} = \eta({\theta}) {\hbar \over 2e} {J\over d}{\bf \hat{m}}\times [{\bf \hat{m}}\times {\bf \hat{M}}],
    \label{eqn:SpinTransfer}
\end{equation}
where ${\bf{\hat M}}={\bf M}/M_s$ and $\cos\theta = \bf{\hat m}\cdot \bf{\hat M}$.

\begin{figure}
    \centering
    \includegraphics[width= 2.75in]{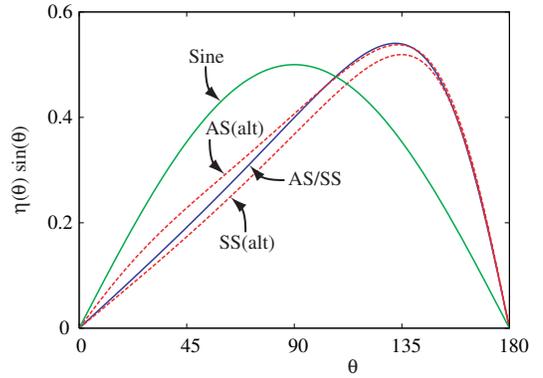}
    \caption{Various forms of (dimensionless) spin-transfer torque as a function of the angle $\theta$ between the fixed layer and the free layer. The sine torque does not depend on the spin valve geometry. The symmetric Slonczewski (SS) and asymmetric Slonczewski (AS) torques are essentially identical for the standard spin valve geometry studied in this paper (AS/SS solid curve). The dashed curves show the difference between the symmetric and asymmetric Slonczewski torques for a geometry discussed in Sec.~\ref{ASSS}.}
    \label{fig:torque}
\end{figure}

The different forms of spin-transfer torque one finds in the literature correspond to different choices for $\eta(\theta)$. If one simply puts $\eta(\theta)=\eta_0$, the result is a ``sine'' approximation to the torque because the remaining angular factors in Eq.~(\ref{eqn:SpinTransfer}) give ${\bf N}_{\rm st} \propto \sin\theta$ (Fig.~\ref{fig:torque}). This form of the torque arises when there is spin-dependent scattering at the free layer interface and the polarization of the electron current that flows from the fixed layer to the free layer is independent of the orientation of the free layer. The prefactor $\eta(\theta)$ is not constant if there is a diffusive component  to the current anywhere and/or spin-dependent reflection occurs at the fixed-layer interface. To our knowledge, one or both of these effects is present in all
transport theory calculations of ${\bf N}_{\rm st}$. On the other hand, the corresponding $\sin^2(\theta/2)$ approximation for the angular dependence of the magnetoresistance describes real spin valve data\cite{Pratt:private,Urazhdin:tbp} better than one would expect based on the transport theory predictions, to which we turn next.

Building on his original work, \cite{Slonczewski:1996} Slonczewski\cite{Slonczewski:2002} applied an approximate form of magnetoelectronic circuit theory\cite{Brataas:2001} to a spin valve with equal lead lengths and equal ferromagnetic layer thicknesses. He found
\begin{equation}
    \eta(\theta) = {q\over A+B\cos\theta},
    \label{eqn:etaTypeI}
\end{equation}
where $q$, $A$, and $B$ are material and geometric parameters. We will call this the symmetric Slonczewski (SS) approximation for the torque.
For a general spin valve geometry, it turns out that\cite{Kovalev:2002,Xiao:2004}
\begin{equation}
    \eta(\theta) = {q_+\over A+B\cos\theta}+{q_-\over A-B\cos\theta}.
    \label{eqn:etaTypeII}
\end{equation}

The present authors have shown that Eq.~(\ref{eqn:etaTypeII})  gives quantitative agreement with calculations of the spin-transfer torque based on the Boltzmann transport equation for a wide variety of spin valve geometries.\cite{Xiao:2004} We will call this the asymmetric Slonczewski (AS) approximation. One of the solid curves in Fig.~\ref{fig:torque} shows that the symmetric and asymmetric  Slonczewski torques are essentially identical for the particular spin valve geometry we use to model the experimental sample of Ref.~\onlinecite{Kiselev:2003} (see Sec.~\ref{minimal}). The two dashed curves show the difference between the the symmetric and asymmetric Slonczewski torques for a spin valve geometry we will discuss in Sec.~\ref{ASSS}.

Spin-transfer torque accounts for non-equilibrium processes that cannot be described by an energy functional. This means that ${\bf N}_{\rm st}$ does not produce an effective field like Eq.~(\ref{eqn:E_total2}) and no damping of spin-transfer dynamics occurs if ${\bf H}_{\rm eff}=0$ and the Landau-Lifshitz form Eq.~(\ref{eqn:LLDamping}) is used for damping.  On the other hand, if one believes that it must be possible to influence spin-transfer driven motion by transferring energy to other degrees of freedom, it is necessary to the use the Gilbert form of damping in the magnetization equation of  motion.  This is what we do in Eq.~(\ref{eqn:Gilbert}).

\subsection{Current-Induced Effective Field}
\label{ef}
First principles calculations\cite{Xia:2002,Stiles:2002b} show that the absorption of a transverse spin current at a ferromagnetic interface is not 100\% efficient. Part of the fraction that survives gives a small correction to $\eta(\theta)$ in Eq.~(\ref{eqn:SpinTransfer}). The remainder is polarized perpendicular to both ${\bf m}$ and ${\bf M}$   and contributes a torque density on the free layer of the form
\begin{equation}
    {\bf N}_{\rm eff} = \eta(\theta)\beta{\hbar \over 2e} {J\over d}{\bf \hat{m}}\times {\bf \hat{M}}.
    \label{eqn:CIEF}
\end{equation}
In the circuit theory language of Brataas {\it et al.}\cite{Brataas:2000} this term is described by the imaginary part of the mixing conductance. Evidently, ${\bf N}_{\rm eff}$ produces motion of ${\bf \hat m}$ identical to that produced by an effective external field oriented along the magnetization direction ${\bf \hat M}$ of the fixed layer. This contribution is usually  neglected because the cited calculations find  $\beta \approx 0.05$. We include it here because at least one experiment\cite{Zimmler:2004} has been interpreted as demonstrating that $\beta \approx 0.20$.
\subsection{Spin Pumping}
\label{sp}
A final contribution to the torque on the free layer comes from a phenomenon called ``spin pumping''. Since a spin polarized current incident from a non-magnet can produce magnetization dynamics in an adjacent ferromagnet, it is not unreasonable that motion of the magnetization of a ferromagnet can influence the spin current in an adjacent non-magnet. The most prominent effect is the injection of a spin current into the non-magnet whenever the magnetization moves.  One consequence of the injected spin current is a back-reaction torque that increases the damping of the spin motion.\cite{Tserkovnyak:2002,Simanek:2003,Mills:2003} This effect has been confirmed by experiments.\cite{Urban:2001,Mizukami:2002,Ingvarsson:2002,Lenz:2004} The torque density due to spin-pumping is given by Tserkovnyak {\it et al.}\cite{Tserkovnyak:2005} as
\begin{equation}
    {\bf N}_{\rm sp} = {1 \over d}{\bf \hat{m}}\times {\bf J}_s^{\rm exch}\times {\bf \hat{m}},
    \label{eqn:Gilbert-spinpumping}
\end{equation}
where
\begin{equation}
    {\bf J}_s^{\rm exch}
    = \half\midb{{\bf J}_s^{\rm sp}-
    \nu({\bf J}_s^{\rm sp}\cdot{\bf \hat{M}})
    {{\bf \hat{M}} - \nu{\bf \hat{m}}\cos\theta \over 1-\nu^2\cos^2\theta}}
\end{equation}
with
\begin{equation}
    {\bf J}_s^{\rm sp}
    = {\hbar g^{\up\down}\over 4\pi S}{\bf \hat{m}}\times{d{\bf \hat{m}}\over dt}.
\end{equation}
$S$ is the cross-sectional area of the free layer. Table I gives numerical values for the parameters $\nu$ and $g^{\up\down}$ (defined in Ref.~\onlinecite{Tserkovnyak:2003}) for the Co/Cu/Co spin valve of interest to us here.

\section{Minimal Model}
\label{minimal}
This section compares LLG simulation results with the experimental results reported in  Ref.~\onlinecite{Kiselev:2003}. Our ``minimal'' model is  Eq.~(\ref{eqn:Gilbert}) with ${\bf N}={\bf N}_{\rm st}$ from Eq.~(\ref{eqn:SpinTransfer}) and the asymmetric Slonczewski (AS) choice in Eq.~(\ref{eqn:etaTypeII}) for $\eta(\theta)$. This model takes account of magnetostatic and surface anisotropy, an external magnetic field, current-induced spin-transfer torque, Gilbert damping, and thermal fluctuations.  Most of our calculations use a spin valve geometry (see Fig.~\ref{fig:spinvalve}) designed to mimic the  nanopillar samples studied by Kiselev {\it et al.}\cite{Kiselev:2003}:
\[
\rm Cu(80~nm)/Co(40~nm)/Cu(thin)/Co(3~nm)/Cu(10~nm).
\]
The notation Cu(thin) indicates that the thickness of the spacer layer is immaterial as long as it is smaller than the mean free path in copper. The precise choice of lead lengths is subject to uncertainty due to the approximations needed to model finite width and reservoir effects in a one-dimensional Boltzmann equation calculation of spin valve transport.\cite{Xiao:2004}

\subsection{Computational Details}
\label{CD}
We solve the stochastic LLG equation using the Ito calculus\cite{Berkov:2002} and a numerical method described by Milshtein.\cite{Milshtein:1978} The simulations proceed by fixing the external field $H$ and sweeping the current density $J$ in steps of size $\delta J$. Before changing to the next value of $J$, we integrate the LLG equation for a ``waiting time'' $t^\ast$ using $N$ time steps of length $\delta t$.  After each time step, we use the instantaneous value of the angle $\theta$ between ${\bf M}$ and ${\bf m}$ and the results of Ref.~\onlinecite{Stiles:2002a} to evaluate the instantaneous magnetoresistance $R(\theta)$.  A time-average over these $N$ values gives the resistance we report for each $J$.

\begin{figure}
   \centering
   \includegraphics[width= 2.75in]{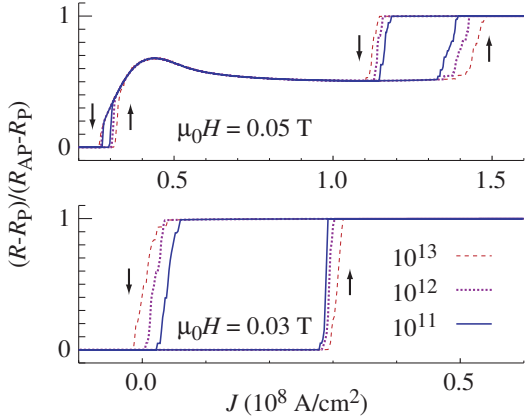}
   \caption{High field (upper panel) and low field (lower panel) magnetoresistance as a function of current density sweep rate (SR) in units of $\rm A/(cm^2$$\cdot$$\rm s)$. The up arrows identity the parts of the hysteresis loops traced out when $J$ is scanned from negative values to positive values. The down arrows correspond to scanning from positive values to negative values of $J$.}
   \label{fig:TendConverge}
\end{figure}

Fig.~\ref{fig:TendConverge} shows the calculated high-field and low-field magnetoresistance as a function of $J$ for three values of the simulated sweep rate ${\rm SR} = \delta J/t^\ast$.  The curves in this figure are averages over 20 realizations of the stochastic simulation.  In each realization, the system switches abruptly at a particular value of current between states with distinctly different magnetoresistance (to be discussed below). Since the switching current depends on the realization, an average over essentially vertical transitions at slightly different switching currents gives the not-quite-vertical lines seen in the figure. As expected, the hysteresis loops close as the sweep rate decreases. Less obviously, the rate of closing is much greater at high field than at low field. It is important to appreciate that the slowest sweep rate we can practically use in our simulations [$\rm 10^{11} A/(cm^2$$\cdot$$\rm s)$] is still five orders of magnitude faster than the sweep rate used in the Cornell experiments.\cite{sweeprate}


For fixed values of $H$ and $J$, the $N$ values of resistance collected between $t=0$ and $t=t^\ast$ constitute a time series for the resistance. Spin valves are Ohmic devices, so the Fourier transform of this series is proportional to the associated power spectrum. We use this numerical data below to compare with the microwave noise data reported in Ref.~\onlinecite{Kiselev:2003}.

\begin{figure*}
\centering \includegraphics[width= 6.5in]{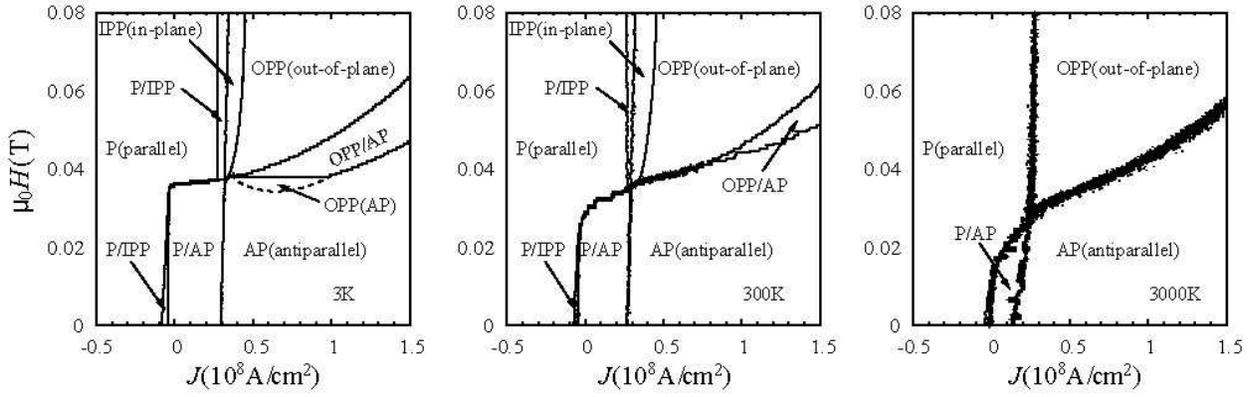}
    \caption{Minimal model dynamic phase diagrams for a Cu(80~nm)/Co(40~nm)/Cu(thin)/Co(3~nm)/Cu(10~nm) spin valve. Left panel: $T=3~{\rm K}$; middle panel: $T=300~{\rm K}$; Right panel: $T=3000~{\rm K}$. For fixed $H$, a bistable region labelled A/B exhibits the A state when $J$ is scanned from left to right and the B state when $J$ is scanned from right to left.  The correspondence needed to compare with Ref.~\onlinecite{Kiselev:2003} is $10^8~{\rm A/cm^2} \leftrightarrow 10~{\rm mA}$. The dashed curve is the OPP$\rightarrow$AP phase boundary for a field scan from large $H$ to small $H$ at fixed $J$.}
    \label{fig:PD-contour}
\end{figure*}

\subsection{$J$-$H$ Phase Diagrams}
\label{HJPhase}
Fig.~\ref{fig:PD-contour} compares spin valve ``phase diagrams'' at $T=3~{\rm K}$, $T=300~{\rm K}$, and $T=3000~{\rm K}$ for our minimal model.  The diagrams were constructed by sweeping the current twice (once increasing the current and once decreasing the current) for each value of $H$. There is some noise at  higher temperature because we did not average over multiple realizations of the simulation. Solid lines divide each diagram into phase fields with labels like A, B and A/B. The latter means that the field is occupied by phase A when the current is scanned from left-to-right in the diagram and by phase B when the current is scanned from right-to-left. Thus, a label like A/B is a signal that hysteresis is present.
\begin{figure}
   \centering
   \includegraphics[width= 2.75in]{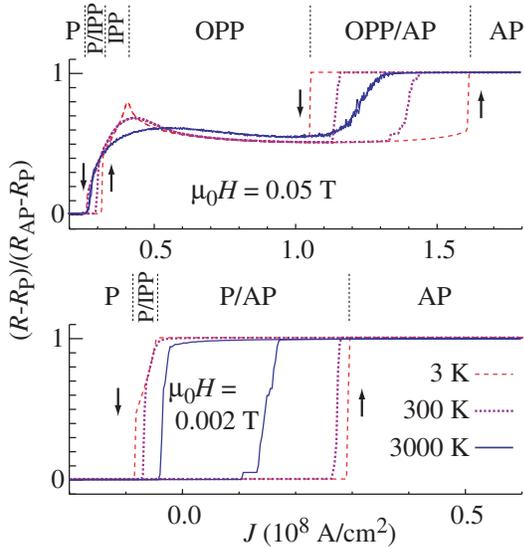}
   \caption{Temperature dependence of the magnetoresistance obtained by averaging the results of 20 realizations of the stochastic simulation. Upper panel: $\mu_0 H=0.05~{\rm T}$; Lower panel: $\mu_0 H=0.002~{\rm T}$.}
   \label{fig:RvsI}
\end{figure}

The phase fields in Fig.~\ref{fig:PD-contour} are labelled P (parallel), AP (anti-parallel), IPP (in-plane precession) and OPP (out-of-plane precession). The static P and AP states are labelled by the relative orientation of ${\bf m}$ and ${\bf M}$.  The precessing states are identified from the microwave power (not shown) as described above.  IPP denotes a dynamic state where ${\bf m}$ precesses symmetrically (or nearly so) around an axis that lies in the $y$-$z$ easy plane. OPP denotes a dynamic state where ${\bf m}$ precesses symmetrically (or nearly so) around an axis that does not lie in the easy plane. Sec.~\ref{orbit} describes these states in more detail.

We focus first on the $3~{\rm K}$ diagram. This is similar (but not identical) to $T=0$ K diagrams published by others\cite{Grollier:2003,Kiselev:2003} using the symmetric Slonczewski spin-transfer torque. Using sharp peaks in the measured noise power spectrum to identify states of stable  precession, Kiselev {\it et al.} pointed out the topological similarity between their computed  $T=0$ K phase diagram and their measured $T=300$ phase diagram.\cite{Kiselev:2003}

When $H$ exceeds the coercive field, our 3 K phase diagram shows hysteresis for the P$\leftrightarrow$IPP and OPP$\leftrightarrow$AP phase transitions. The experiment shows no hysteresis in this regime (see below). At low field, the $\rm P\to AP$ transition occurs abruptly while the reverse-current $\rm AP\to P$ transition does not. Instead, there is a long, skinny, triangular-shaped P/IPP phase field within which the magnetization ${\bf m}$ exhibits stable, elliptical precession around the $-\hat{\bf z}$ axis.  The precession amplitude increases as the current becomes more negative. The system crosses the phase boundary into the P phase when the precession angle between ${\bf m}$ and $-{\bf \hat{z}}$ exceeds 90$^{\circ}$ and the vector ${\bf m}$ spirals irreversibly toward ${\bf \hat{z}}$. We will see below that this asymmetry has its origin in the details of the dependence of the spin-transfer torque on the angle between the free layer and the fixed layer.

We draw special attention to the lower limit of the OPP/AP phase field in the 3 K phase diagram. The perfectly horizontal portion of this phase boundary is an artifact of the current scanning mode used to generate the diagram. If we fix $J$ and scan the external field $H$ from large values to small values, the OPP phase does not give way to the AP phase until the dashed line in the diagram is crossed. The exact shape of this boundary depends on the $H$-scan rate, but it is reasonable to suppose that the corresponding horizontal phase boundary found in Ref.~\onlinecite{Kiselev:2003} may also be an artifact of the method of taking data. 

Reading Fig.~\ref{fig:PD-contour} from left to right shows that the regions of hysteresis shrink as the temperature increases. More details can be seen in the line scans of Fig.~\ref{fig:RvsI}.  The effect of increasing temperature is very similar to the effect of decreasing the current density sweep rate in Fig.~\ref{fig:TendConverge}.  Indeed, since our simulated current sweep rate is always much faster than experiment, the temperatures indicated on the phase diagrams in Figure~\ref{fig:PD-contour} must be regarded as nominal.  The true phase diagram at each temperature we show would exhibit less hysteresis. Equivalently, each panel actually corresponds to a lower physical temperature than the temperature we quote. Thus, our $T=$~3000~K  diagram indicates (qualitatively) how the 300~K phase diagram might look if we could use current sweep rates comparable to those used experimentally. We note also that substantial Joule heating occurs in real spin valve samples, perhaps 15~K to 20~K per 10$^{\rm 7}$ A/cm$^{\rm 2}$.\cite{Krivorotov:2004}

\begin{figure}
  \centering \includegraphics[width= 2.75in]{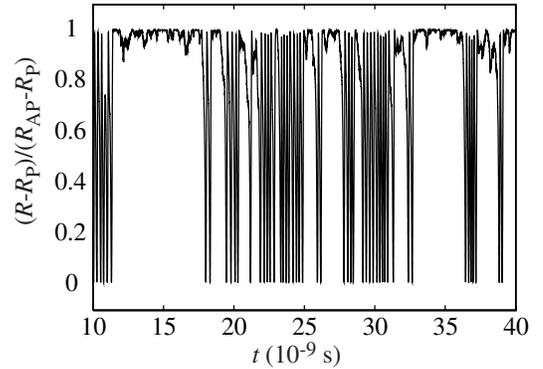}
  \caption{Telegraph noise in a time-series of the magnetoresistance. The data was collected on the high-temperature boundary between the AP phase and the OPP phase ($\mu_0 H=0.04$T, $J=0.7\times 10^8 {\rm A/cm^2}$).}
  \label{fig:tele}
\end{figure}

Our highest temperature simulation shows P$\leftrightarrow$AP hysteresis when $H$ is small and complete reversibility when $H$ is large. This resolves the disagreement between theory and experiment noted above.  Moreover, state-to-state switching characterizes every reversible phase boundary. Fig.~\ref{fig:tele} illustrates this for the dynamics of switching between the anti-parallel AP state and out-of-plane precession (OPP). The time-series data for the magnetoresistance was collected on the OPP/AP phase boundary at 3000 K. Clearly, the system switches back and forth between the AP state (small fluctuations around unit normalized resistance) and the OPP state (periodic oscillations of the normalized resistance between zero and one).  Experiments show precisely this sort of telegraph noise\cite{Myers:2002,Urazhdin:2003,Pufall:2004,Krivorotov:2004} if we replace the full scale excursions of the OPP resistance with small scale fluctuations around the average resistance of the OPP state.
We are not aware of experiments that study the telegraph noise at our 300 K IPP/OPP boundary or our 3000 K P/OPP boundary. Indeed, at the latter, our simulations actually show random switching between {\it three} states: AP,IPP, and OPP.

The variations of the computed resistance near the 3000~K P/OPP and OPP/AP phase boundaries lead to two peaks in the differential resistance, $dV/dI=R+I\ dR/dI$. These agree well with the peaks in $dV/dI$ observed experimentally. On the other hand, Kiselev {\it et al.}\cite{Kiselev:2003} identify a ``W''-phase that is completely absent from our 3000~K phase diagram. However, the experimental W-phase field appears exactly where our model predicts OPP/AP phase bistability at 3 K and 300 K (two left panels of Fig.~\ref{fig:PD-contour}). The experimental W-phase is microwave quiet above the experimental low frequency cut-off (0.1~gHz) and it exhibits a magnetoresistance that is slightly, but distinctly, smaller than that of the AP configuration. This would occur in our macrospin model if the free layer were frozen into a static configuration with ${\bf m}$ neither parallel nor anti-parallel to ${\bf M}$.  We will return to the W-phase when we discuss micromagnetic simulations in Section~\ref{micromag}.

Quantitatively, our calculated coercive field is about half the experimental value. This discrepancy may reflect an inaccurate description of the shape (and therefore the magnetostatic anisotropies) of the free layer. Another contributory effect is our complete neglect of dipolar coupling to the fixed layer. At low $T$, we also find that the magnitude of the critical current $J_c^+$ for the P$\rightarrow$AP transition  is much greater than the magnitude of the critical current $J_c^-$ for the AP$\rightarrow$P transition. Experiments show that $J_c^+$ and $J_c^-$ are more symmetric around zero current. The reason for our calculated behavior is\cite{Slonczewski:1996} the difference between $\eta(0)$ and $\eta(180^\circ)$ for the Slonczewski torques in Fig.~\ref{fig:torque}.  Our simulations at 3000 K more nearly resemble the experiments at 300 K because $J_c^+$ decreases strongly with temperature while $J_c^-$ is nearly temperature independent.

\subsection{Precession Frequency}
\begin{figure}
    \centering
    \includegraphics[width= 2.75in]{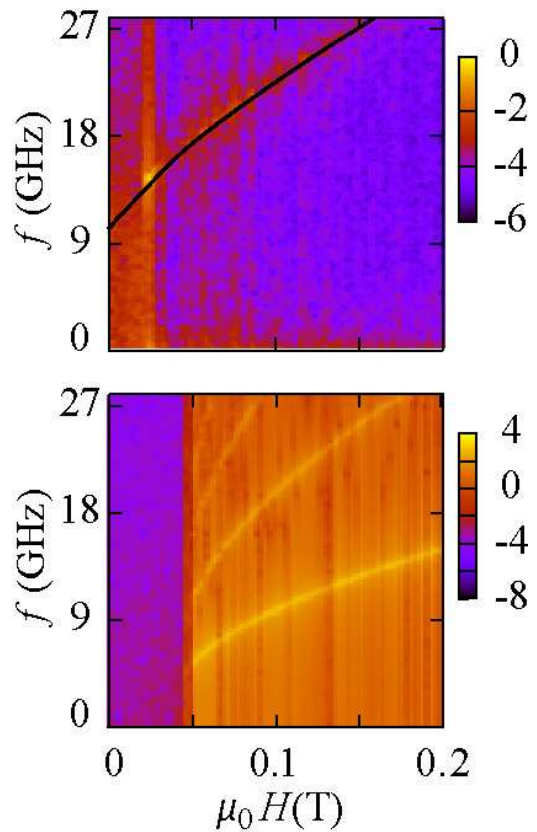}
\caption{Relative microwave power at different frequencies as a function of field $H$ at $T=3~K$. The gray/color scale is logarithmic. Left panel: $J=0.3\times 10^8{\rm A/cm^2}$;
the black  curve is twice the resonance frequency given by the
Kittel equation Eq.~(\ref{eqn:kittel_freq}). Right panel: $J=1.0\times 10^8{\rm A/cm^2}$. The multiple curves are the fundamental and its harmonics.}
    \label{fig:freq_field}
\end{figure}

 The gray/color scale in
Fig.~\ref{fig:freq_field} quantifies the relative microwave power (on a logarithmic scale)
at frequency $f$ as a function of magnetic field for two values of current density $J$. The numerical data was obtained by Fourier transforming our simulated time series data for the magnetoresistance. The narrow bands of peak microwave power trace out the frequency $\omega(H)$ of stable precession (and its harmonics). The left panel corresponds to small-amplitude, noise-driven, in-plane precession at a value of current just before the parallel phase becomes unstable to steady in-plane-precession. As with the experimental data,\cite{Kiselev:2003,Rippard:2004a} $\omega(H)$ in this regime can be described by the Kittel equation (black curve) for thin film magnetic resonance.\cite{Kittel:1948}  In the notation of Appendix I, the resonance frequency is
\begin{equation}
    \omega_K
= \gamma\sqrt{[H+(h_Y-h_Z)M_{\rm s}][H+(h_X-h_Z)M_{\rm s}]}.
    \label{eqn:kittel_freq}
\end{equation}
Our simulation data agree with $2\omega_K$ because the periodicity of the resistance
is twice the periodicity of the magnetization oscillation frequency. No analytic theory is available for comparison with our results at a higher value of $J$ (right panel) where large amplitude out-of-plane precession occurs. But our results do show the same relative magnitude and $H$-dependence as seen in the experiments.

\begin{figure}
    \centering
    \includegraphics[width= 2.75in]{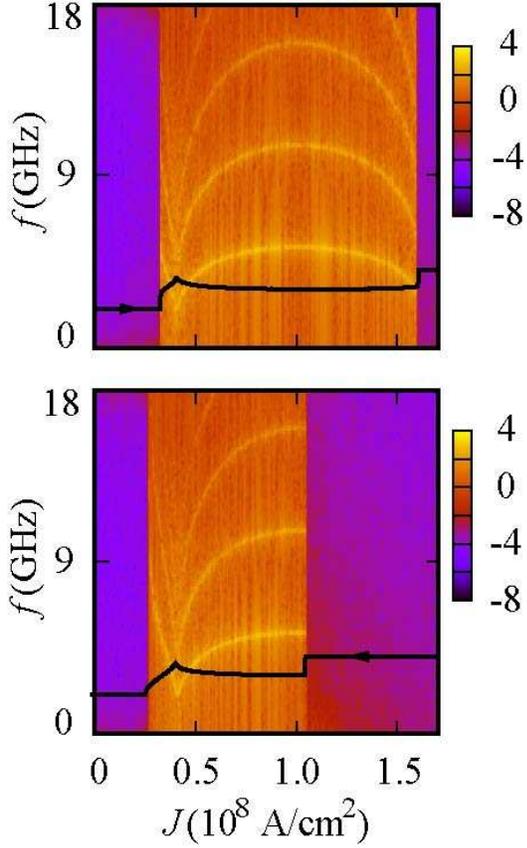}
    \caption{Microwave power at different frequencies as a function of $J$ at $T=3~{\rm K}$ and $\mu_0 H=0.05~{\rm T}$. The gray/color scale is logarithmic. Left panel: current scan from left to right. Right panel: current scan from right to left. The narrow bands of peak power represent the frequency $\omega(J)$ of stable precession. The black traces are the magnetoresistance in arbitrary units.}
    \label{fig:freq_default}
\end{figure}

Fig.~\ref{fig:freq_default} shows the relative microwave power at frequency $f$ as a function of increasing $J$  (top panel) and decreasing $J$ (bottom panel). Similar plots for comparison with experiment have been presented by others using the symmetric Slonczewski torque\cite{Kiselev:2003} and the sine torque.\cite{Russek:2005} The zero power regions at low and high $J$ correspond to the static P and AP magnetization states. In between, the narrow bands of peak microwave power trace out $\omega(J)$ (and its harmonics) for stable precession. Just above the limit of the parallel state, there is a very narrow range of in-plane precession  where $\omega(J)$ decreases monotonically.  At slightly higher $J$, the system evolves to a state of out-of-plane precession (OPP) where $\omega(J)$ first increases and then decreases. Comparison of the two panels in Figure~\ref{fig:freq_default} illustrates the hysteresis present at this low temperature.

Our results for $\omega(J)$ do not agree with observations for real spin valves.\cite{Kiselev:2003,Rippard:2004a} Putting aside the fact that no hysteresis is seen in the experiments (which we attribute to the current sweep rate as discussed above), the experimental data always show that $\omega$ decreases as $J$ increases. Naively, it is as if in-plane precession persisted all the way to the anti-parallel state with no intervening state of out-of-plane precession.  This is a serious issue because, in our model, in-plane precession occupies an extremely small portion of the $J$-$H$ phase diagram.

\subsection{Precession Trajectories}
\label{orbit}
To help shed light on our simulation results for $\omega(J)$, it is instructive to analyze the relationship between this quantity and the trajectory of the tip of ${\bf \hat{m}}$ on the unit sphere. We will call this the orbit of the precessional motion. Without loss of generality, we set $h_Y = h_Z = 0$ and retain only the external field ${\bf H}=H\hat{\bf z}$ and the {\it local}, out-of-plane demagnetization field ${\bf H}_d= -h_X \hat{m}_x \hat{\bf x}$. It is crucial that the  magnitude $H_d$ changes along the trajectory because the component $\hat{m}_x$ of ${\bf \hat{m}}$ changes along the trajectory.
\begin{figure}
    \centering
    \includegraphics[width= 1.75in]{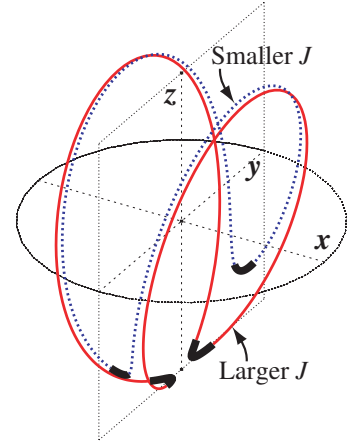}
    \caption{In-plane precession (IPP) orbits for two nearby values of $J$.    The thick segments are points on each orbit where the magnitude of the demagnetization field ${\bf H}_d= -h_X \hat{m}_x \hat{\bf x}$ is smaller than the magnitude of the external field ${\bf H} = H {\bf \hat{z}}$. The $y$-$z$ easy plane and the equatorial circle of the  unit sphere in the $x$-$y$ plane are indicated as guides to the eye.}
    \label{fig:ipp}
\end{figure}

\begin{figure}
    \centering
      \includegraphics[width= 1.75in]{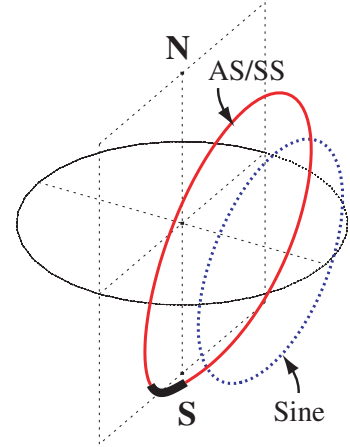}
    \caption{Out-of-plane precession (OPP) orbits. The geometry is the same as Fig.~\ref{fig:ipp} except that the north (N) and south (S) poles of the unit sphere are indicated. The orbit labelled ``AS/SS'' is produced by Slonczewski's spin-transfer torque. Along the thick segment, the spin-transfer torque and demagnetization field torque point in opposite directions. The orbit labelled ``sine'' is produced by a sine-type spin-transfer torque for the same value of $J$.}
    \label{fig:opp}
\end{figure}

It is common to think of precession as the steady motion of a vector on a cone that makes a small angle with respect to its symmetry axis. The orbit in this case is a circle. The precessional states in the present problem are more complicated.  Fig.~\ref{fig:ipp} shows two large amplitude, saddle-shaped, in-plane precession (IPP) orbits for two nearby  values of $J$.  We call these ``in-plane'' precession modes because each orbit moves symmetrically (or nearly so) around an axis (the $z$-axis) that lies in the easy $y$-$z$ plane.

Let us partition each orbit into two segments. The short thick segments lie near the easy plane where the demagnetization field $H_d$ is smaller than the external field $H$. Along these segments, the orbital azimuthal angle $\phi$ precesses mainly around ${\bf H}$ with angular speed  $\gamma H$. Along the remaining segment of each orbit, $H_d$ is larger than $H$ and the orbital polar angle $\theta$ precesses  mainly around ${\bf H}_d$ with angular speed $\gamma H_d$. Fig.~\ref{fig:ipp} shows that the angular range swept out by {\it both} the thick and thin segments increases as the current density (and the spin-transfer torque) increases, {\it i.e.}, the total arc length of the orbit increases. Since the orbital speeds change very little with $J$, we conclude that the orbital period increases as current density increases. This implies that $\omega(J)$ is a decreasing function for in-plane-precession orbits.

As $J$ continues to increase, the apices of the two thick segments of the in-plane precession saddle orbit approach and then touch one another near the negative $z$-axis. When this occurs the orbit bifurcates into two elliptical orbits, each centered on an out-of-plane axis not far from the $x$-axis.\cite{Kiselev:2003,Morise:2005,Russek:2005,StilesMiltat} Precessional states at higher current density correspond to one or the other of these out-of-plane (OPP) trajectories, {\it e.g.}, the AS/SS orbit in  Fig.~\ref{fig:opp}. This orbit precesses mostly around ${\bf H}_d$.  Spin-transfer torque tends to push the orbit away from the easy plane in the northern unit hemisphere. The effect on the orbit in the southern unit hemisphere is more complex.  The net result is that the ``center'' of the orbit moves away from the easy plane. In other words, as the current density increases,  the component $\hat{m}_x$ of ${\bf \hat{m}}$ increases, which increases $H_d$, and thus increases the frequency $\gamma H_d$ of the orbit.

Along the thick segment of the out-of-plane orbit, the spin-transfer torque and the torque from the demagnetization field point in (nearly) opposite directions. This means that the net torque, and thus the orbital speed along that segment, decreases as $J$ increases. Eventually, this slowing down overwhelms the speeding up described just above and the total orbital period begins to increase. This is why $\omega(J)$ decreases for the largest values of $J$ where precession occurs in the top panel of Fig.~\ref{fig:freq_default}.

\section{Beyond the Minimal Model}
\begin{figure*}
  \centering
  \includegraphics[width= 6.5in]{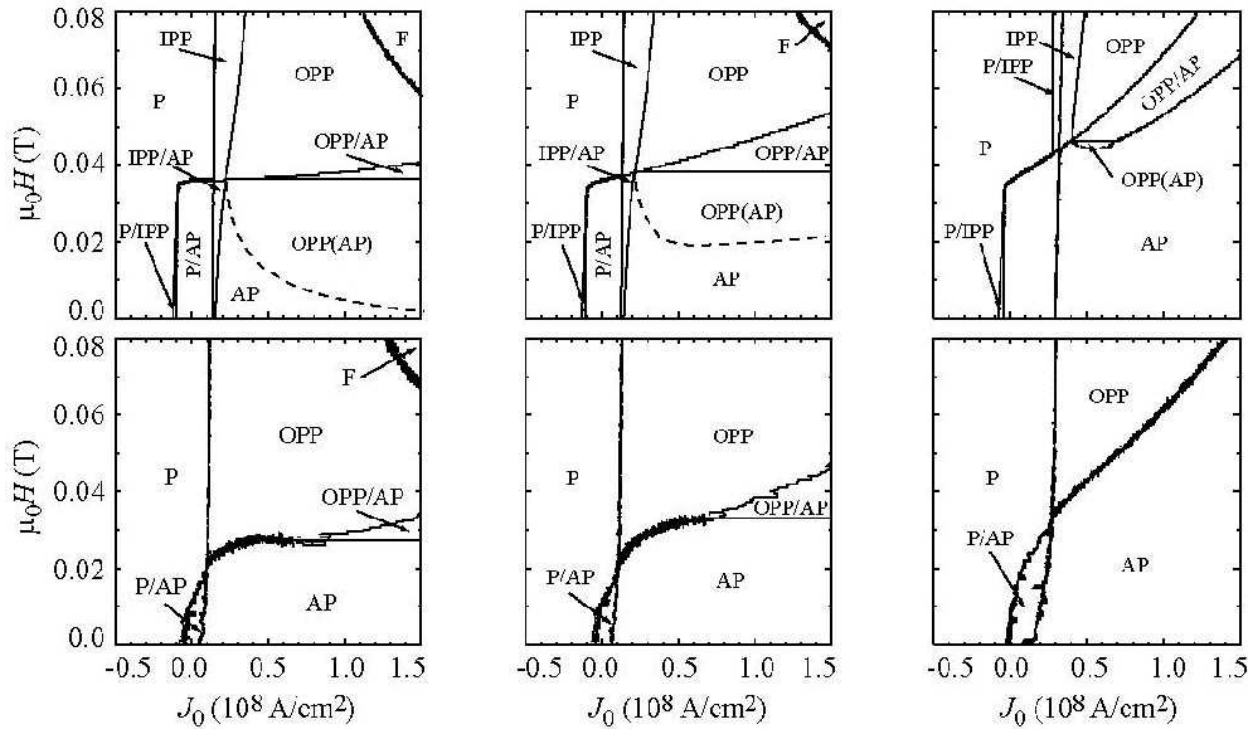}
\caption{Phase Diagrams at 3 K (top panels) and 3000 K (bottom panels). Left panel:  sine torque; middle panel: sine torque plus current-induced effective field; right panel: Slonczewski torque plus current-induced effective field.  For fixed $H$, a bistable region labelled A/B exhibits the A state when $J$ is scanned from left to right and the B state when $J$ is scanned from right to left. The dashed curves are the OPP$\rightarrow$AP phase boundaries for a field scan from large $H$ to small $H$ at fixed $J$.}
  \label{fig:as_ss_sine_EF}
\end{figure*}

There are two major discrepancies between the Cornell experiment\cite{Kiselev:2003} and our minimal model results:  the variation of the precession frequency $\omega$ with current density $J$ and the absence of a microwave quiet ``W-phase''. Within the context of the macrospin model, we examined several variations of our model, mostly with the hope they would improve the agreement between theory and experiment. We studied the influence of (A) a sine-type spin transfer torque for the standard geometry; (B) asymmetric vs. symmetric Slonczewski spin-transfer torque for a special  asymmetric geometry; (C) the current-induced effective field that arises due to incomplete absorption of transverse spin currents; (D) spin-pumping;  and (E) angle-dependent Gilbert damping.
\subsection{Sine Spin-Transfer Torque}
\label{other}

 Fig.~\ref{fig:torque} shows the geometry-independent ``sine'' torque that is widely used in the literature. The top left panel in Fig.~\ref{fig:as_ss_sine_EF}, shows the 3 K phase diagram when this sine torque replaces the AS/SS torque. Several differences with the corresponding Slonczewski torque phase diagram (left panel of Fig.~\ref{fig:PD-contour}) should be noted.

First, with a sine torque, the low-field P$\rightarrow$AP transition is mediated by in-plane precession in the same way that precession mediates the AP$\rightarrow$P transition for both the sine and Slonczewski torques.
This occurs because $\sin\theta$ is symmetric around $\theta=\pi/2$ while the minimal model torque is not. Second, the sine torque generates no hysteresis in the high-field transitions  P$\leftrightarrow$IPP and IPP$\leftrightarrow$OPP. Third, the lower limit of the  OPP phase boundary determined by a field scan from large $H$ to small $H$ (dashed curve) greatly reduces the size of the AP phase field compared to the Slonczewski case. This feature does not appear to have been noticed in previous discussions of this phase diagram.\cite{Xi:2004,Morise:2005} 

Unlike the Slonczewski torque, increasing current or field eventually drives the sine torque model to a transition from out-of-plane precession (OPP) to a  static  phase where the macrospin ${\bf m}$ is ``fixed'' (F) at some angle between zero and $\pi$. This is intriguing because the magnetoresistance and microwave power characteristics of this phase match exactly to those of the experimentally observed ``W-phase''. Unfortunately, the location of the F phase in the sine torque phase diagram does not agree with the location of the W-phase in the experimental phase diagram (see the penultimate paragraph of Sec.~\ref{HJPhase})

The bottom left panel in Fig.~\ref{fig:as_ss_sine_EF} shows the 3000 K phase diagram for the sine torque macrospin model. Compared to the corresponding diagram for the minimal model (right panel of Fig.~\ref{fig:PD-contour}), thermal effects eliminate OPP/AP bistability only when $J$ is small.  Hysteresis between these phases remains when $J$ is large. In this sense, the sine torque model is more resistant to thermal fluctuations than the Slonczewski torque model. 

Finally, the out-of-plane precession frequency for the sine torque is a strictly increasing function of $J$. The argument is similar to the one in Sec.~\ref{orbit} for the Slonczewski torque. However, as the current increases, the different angular dependence of the sine torque causes the orbit of ${\bf \hat{m}}$ to push steadily away from the easy plane everywhere and contract on the unit sphere (see Fig.~\ref{fig:opp}). The frequency $\gamma H_d$ increases monotonically because the demagnetization field $H_d$ increases.
The outward motion and areal contraction of the orbit continues as the current increases until the orbit area shrinks to a single point on the unit sphere. This is the signature of the fixed (F) phase.
\subsection{AS vs. SS Spin-Transfer Torque}
\label{ASSS}
\begin{figure}
    \centering
    \includegraphics[scale=0.75]{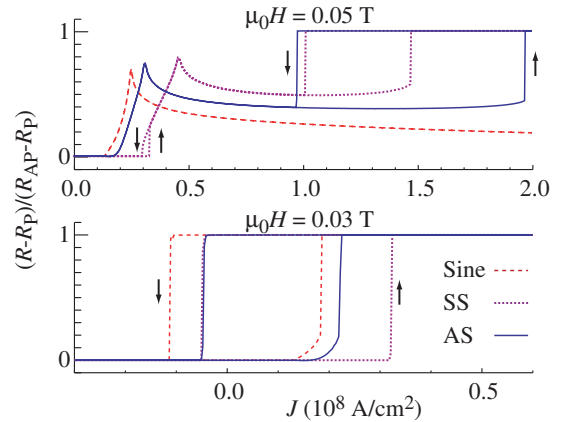}
    \caption{Spin valve hysteresis at 3 K for three choices of spin transfer torque using a geometry (defined in Sec.~\ref{ASSS}) chosen to emphasize the difference between the AS and SS torques. Lower panel: $H=0.03$T; Upper panel: $H=0.05$T.}
    \label{fig:TYPE_I_II}
\end{figure}

The AS/SS curve in Fig.~\ref{fig:torque}  shows that the asymmetric Slonczewski torque used in our minimal model is essentially identical to the symmetric Slonczewski torque for the spin valve geometry of Ref.~\onlinecite{Kiselev:2003}.
This is not always the case. For example, compared to the geometry of Kiselev {\it et al.},\cite{Kiselev:2003} a spin valve with film thicknesses,
\[
\rm Cu(10~nm)/Co(40~nm)/Cu(thin)/Co(3~nm)/Cu(180~nm),
\]
is very asymmetric. The Cu/Co bilayers on opposite sides of the spacer layer are very different: the left  bilayer is mostly ferromagnet, the right bilayer is mostly non-magnet.
The difference between the symmetric and asymmetric Slonczewski torques for this geometry is still small (compare the two dashed curves in Fig.~\ref{fig:torque}). Nevertheless, it is large enough to produce small-angle, in-plane precession that
``rounds'' the low-to-high resistance jump in the AS hysteresis curve at the P$\rightarrow$AP transition in the lower panel of Fig.~\ref{fig:TYPE_I_II}. The same panel shows similar precessional rounding for the sine-type  spin-transfer torque. However, the latter rounding disappears when shape anisotropy is turned off. The corresponding rounding for the asymmetric Slonczewski torque does {\it not} disappear when  shape anisotropy is turned off.\cite{Xiao:2004}

Fig.~\ref{fig:TYPE_I_II} also shows that the critical current $J_c^+$ for the P$\rightarrow$AP transition differs for all three spin-transfer torques while the critical current $J_c^-$ for the reverse AP$\rightarrow$P transition distinguishes only the sine torque. This is a consequence of the fact that $J_c^+$ ($J_c^-$) is inversely proportional the slope of the torque function $\eta(\theta)\sin\theta$ (plotted in Fig.~\ref{fig:torque}) at $\theta=0$ ($\theta=\pi$).\cite{Slonczewski:1996} The fact that $J_c^+$ is smaller for the AS torque than for the SS torque suggests that an asymmetric geometry like the one above may be desirable for some applications.


\subsection{Current-Induced Effective Fields}
We mentioned in Sec.~\ref{ef} that a current-induced torque that acts like an effective external
magnetic field can arise due to incomplete absorption of a transversely polarized spin current at the interface between the spacer and the free layer.  A recent experimental report\cite{Zimmler:2004} has been interpreted
by its authors to mean that the size of this torque---displayed in Eq.~(\ref{eqn:CIEF})--- is much larger than theoretical estimates. Accordingly,  Fig.~\ref{fig:as_ss_sine_EF} shows how the  spin valve 3 K phase diagram changes if we augment the sine torque (top middle panel) and the minimal model torque (top right panel) by an effective field torque that is 20\% of the spin-transfer torque, as suggested by this experiment. For comparison, the bottom middle and bottom right panels respectively show the phase diagrams for these two situations at 3000 K. We find that the topology of the phase diagram does not change, although the precise positions of the phase boundaries do. The qualitative behavior of the precession frequencies is not affected.

\subsection{Spin-Pumping}
The original discussion of spin-pumping\cite{Tserkovnyak:2003} focused on the enhancement of Gilbert damping that occurs when a normal metal is in intimate contact with a precessing thin film ferromagnet. Subsequent work\cite{Tserkovnyak:2005} has emphasized that the torque due to spin-pumping is generally of the same order of magnitude as spin-transfer torques. Since the analytic form of the torque ${\bf N}_{\rm sp}$ in Eq.~(\ref{eqn:Gilbert-spinpumping}) differs considerably from simple damping for, {\it e.g.}, large-angle, out-of-plane precessional motion, this raises the possibility that spin-pumping alters the dynamical behavior of a spin valve more profoundly than merely enhancing the Gilbert damping.

\begin{figure}
   \centering
   \includegraphics[scale=0.8]{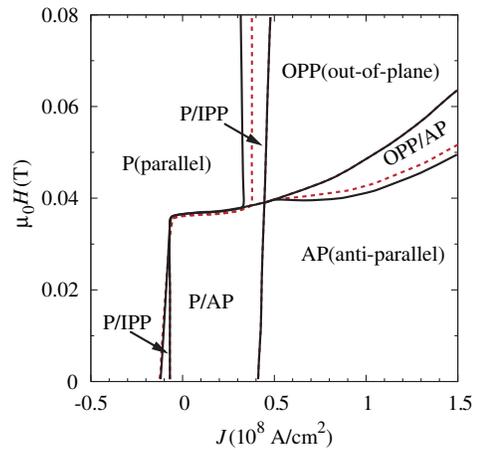}
   \caption{Minimal model phase diagram at 3 K with Gilbert damping only (solid lines) and spin-pumping
   with reduced Gilbert damping (dashed lines). See text for discussion.}
   \label{fig:SP_ED}
\end{figure}

In the small angle limit, the parameters for intrinsic Gilbert damping and spin-pumping suggested in Ref.~\onlinecite{Tserkovnyak:2005} for the Co/Cu/Co system (Table I) produce a total effective Gilbert damping of $\alpha_{\rm eff} = 0.148$. Therefore, in Fig.~\ref{fig:SP_ED}, we compare the phase diagram at 3 K obtained with our minimal model (no spin pumping) using $\alpha = \alpha_{\rm eff}$  (solid lines) with the phase diagram obtained including both (reduced) Gilbert damping and the spin-pumping torque density  ${\bf N}_{\rm sp}$ (dashed lines). The small differences we find between the two show that spin-pumping does not much affect the sort of precessional motion produced by our minimal model. There is also no qualitative change in the current dependence of the precession frequency. We conclude that neither effective fields nor spin pumping affects improves the agreement between experiment and the minimal model.
\subsection{Angle-Dependent Gilbert Damping}
The numerical value of the Gilbert damping constant $\alpha$ in Eq.~\ref{eqn:Gilbert}
is usually extracted from ferromagnetic resonance or Brillouin light scattering experiments.\cite{Mills:2003b}
Since the magnetization is never tilted far from equilibrium in these situations, it is relevant that Back {\it et al.}\cite{Back:1999} reported an effective increase in the damping constant for large magnetization rotation angles in cobalt films under pulsed field conditions. More recent pulsed-field experiments have been successfully analyzed using conventional Gilbert damping. \cite{Kabos:2000,Hiebert:2002} Since transient magnetization dynamics is a common feature of all these experiments, it is not obvious that the results address the validity of the constant $\alpha$ approximation for large-angle, steady precession of the sort discussed in this paper. With one exception,\cite{Katine:2000b} we are unaware of any models that allow the damping constant to vary during the course of precession.

A glance back at Fig.~\ref{fig:freq_default} shows that the precession frequency $\omega(J)$ decreases monotonically (as seen in experiment) for in-plane-precession and also for out-of-plane precession near the boundary with the anti-parallel state. Using the damping ``constant'' {\it ansatz}
\begin{equation}
\label{eqn:AG}
\alpha(\theta)= a+ b\sin^2 \theta,
\end{equation}
we have been able to produce dynamical phase diagrams with (i) only in-plane-precession or (ii) only out-of-plane precession with (in both cases) the ``correct'' behavior for $\omega(J)$, or nearly so.  Unfortunately, other features appear in the simulated phase diagram that do not appear in the data, {\it e.g.}, an extended region of precession when the external field $H$ is less than the coercive field.  It may be possible to fine-tune the phase diagram to the desired form, but this seems unwarranted without some justification for Eq.~(\ref{eqn:AG}) and the relative paucity of theoretical information about damping far from equilibrium.\cite{Dobin:2003}

\section{Micromagnetics}
\label{micromag}
Our macrospin approach to the Landau-Lifshitz-Gilbert (LLG) equation replaces the free layer magnetization ${\bf m}({\bf r})$ by a constant vector $\bf m$.  The numerical method of micromagnetics\cite{Hubert:1998} is a better approximation to reality because it retains spatial gradients of ${\bf m}({\bf r})$ down to a fixed minimum length scale.  Treating the spatial variation of the magnetization allows the inclusion of two effects that we have neglected because they do not contribute when the magnetization is uniform.  First, the experimental samples are polycrystalline so we ignore any intrinsic magnetocrystalline anisotropy because the non-uniform effective fields tend to average to zero over the whole sample. Second, the Oersted magnetic field produced by the current itself largely averages to zero over the whole sample.  While these two effects are not important for the macrospin dynamics we have considered here, full simulations \cite{Berkov:2005b} show that the inhomogeneities in the magnetization that result can be quite important.  

There is an emerging consensus amongst micromagnetic practitioners\cite{Miltat:2001,Li:2003,Zhu:2004,Lee:2004,Montigny:2005,Berkov:2005,Berkov:2005b,Xi:2005} that the inclusion of spin-transfer torque excites incoherent spin waves in the free layer (and thus inhomogeneous magnetization) if the current density is sufficiently great to induce switching and microwave emission.  However,  since the method takes full account of local exchange and non-local magnetostatics, systematic survey calculations of the sort we have presented in this paper are prohibitively expensive, even for systems as small as a spin valve free layer.  
Using a torque density $\rm N _{st} \propto \sin \theta$ to model spin-transfer, Ref.~\onlinecite{Lee:2004} reports finite-temperature micromagnetic simulations designed to mimic the experimental conditions reported by Kiselev {\it et al.}\cite{Kiselev:2003} Like our Fig.~\ref{fig:PD-contour}, the calculated phase diagram agrees  topologically and semi-quantitatively with the experimental diagram, with one important improvement. The micromagnetics simulation identifies the  experimental ``W-phase'' with a dynamical phase field where vortices of magnetization continuously form and annihilate. The calculated noise power in this regime is concentrated at very low frequency and thus appears to be microwave quiet in the experiment. 

It seems plausible that the extra degrees of freedom present in the micromagnetic approximation allow the bistable OPP/AP phase present in our macrospin model to break up into a spatially inhomogeneous state. The same logic suggests that the fixed F phase of the macrospin sine torque phase diagram breaks up similarly into a state of inhomogeneous magnetization. On the other hand, the micromagnetic simulation produces a nearly horizontal line for the AP phase boundary that we identify as an artifact of the current-scanning mode of data collection. 

It is worth noting that the microwave power output in this micromagnetic calculation is nearly independent of $J$ in the high-field precession regime (as we find for a macrospin) while the power output observed experimentally varies considerably in this part of the phase diagram. The precession frequency $\omega(J)$ in the same regime has been studied by Berkov and Gorn\cite{Berkov:2005b}, who find that they are able to qualitatively reproduce the experimental frequency dependence with a highly inhomogeneous magnetic state.  The inhomogeneous state arises from the non-uniform Oersted field and an (assumed) random distribution of granular magnetocrystalline anisotropy.  It would be interesting to discover if this type of micromagnetic simulation can produce resonance linewidths with $Q\approx 100$ as observed in the most recent nanopillar experiments.\cite{Krivorotov:2005}

\section{Summary \& Conclusion}
We have studied the Landau-Lifshitz-Gilbert (LLG) dynamics of a single macrospin as a model for  current-driven magnetization motion in the free layer of a spin valve. We parameterized our model specifically to compare our results with those reported by Kiselev {\it et al.}\cite{Kiselev:2003} for a Cu/Co/Cu nanopillar. Due to the simplicity of the model (compared to micromagnetics), we were able to explore systematically the effects of temperature, spin-pumping, current-induced effective fields, various forms of spin-transfer torque, and angle-dependent damping. We focused most of our attention on a ``minimal model'' where Slonczewski's spin-transfer torque supplements the terms usually found in the LLG equation.

{\it Low-Field Behavior:} Our minimal macrospin model captures the essential features of the experiment when the external field $H$ does not exceed the coercive field of the free layer. As a function of current density $J$, there is hysteretic switching between parallel and anti-parallel orientations of the free layer and the fixed layer. In the experiment, the critical currents for P$\to$AP and AP$\to$P switching have opposite sign but are approximately equal in magnitude . This is not a feature of our $T=3$~K phase diagram, but it is much more nearly true in our simulation at 3000 K, where thermal fluctuations are large enough to mimic the effect of the (slow) current sweep rate used in the experiment. The scale we calculate for $H$ is about half as large as seen in the experiment.

{\it High-Field Behavior:} The macrospin model correctly models noise-driven, low amplitude, in-plane precession  when $H$ is larger than the coercive field and the current density is low. The existence of large-amplitude, in-plane and out-of-plane precession at higher $J$ agrees qualitatively with observation, but the precession frequency function $\omega(J)$ is not monotonically decreasing as found in experiment. The simulation predicts peaks in $dV/dI$ associated with telegraph-noise switching between two (or more) states of magnetization.  These peaks are present in the experiment, as is two-state telegraph noise. We find an OPP/AP bistable phase field that is occupied in the experiment by a microwave quite ``W''-phase. Micromagnetic simulations by others suggest that that vortex creation and annihilation occurs in this phase field.

{\it Other Effects:} The phase diagram of the sine-torque model differs mostly in detail with the phase diagram of the minimal model.  An exception is the presence of a high-field, high current ``fixed'' phase for the sine torque where the macrospin freezes into a fixed angle with respect to the fixed layer. This phase does not seem to occur in micromagnetic simulations. Another exception is the relative persistence of the OPP phase when we scan from high values of $H$ to low values of $H$. This behavior is also absent from the micromagnetics simulations.  Spin-pumping and current-induced magnetic fields do not change the phase diagram in any significant way. Large topologically changes do occur if we allow the Gilbert damping parameter to change with angle, but we have no guidance for the form $\alpha(\theta)$ should take. This seems like a fruitful direction for future research.

\section{Acknowledgment}
 The authors acknowledge useful
discussions about spin valve experiments with Ilya Krivorotov, Bob Buhrman,
Jack Sankey, Bill Rippard, Steve Russek, Tom Silva, Matt Pufall, Sergei Urazhdin, Jack Bass, and Mark Covington.
One of us (J.X.) is grateful for support from the Department of Energy under grant DE-FG02-04ER46170.

\appendix
\begin{center}
    {\bf APPENDIX: ENERGY EXPRESSION}
\end{center}
The energy of the free layer includes a Zeeman energy $E_Z$ from the external field $H$,
a magnetostatic shape anisotropy energy $E_s$, and a surface anisotropy energy (parameterized by $K_u$) that vanishes in the limit
that the free layer thickness $d\to\infty$. 

The Zeeman energy is
\begin{equation}
    E_Z = V  \mu_0 M_{\rm s} \hat{\bf m}\cdot{\bf H}. \vspace{0.5em}
    \label{eqn:E_zeeman}
\end{equation}
The shape anisotropy energy is
\begin{equation}
    E_s = V\half \mu_0 M_{\rm s} \hat{\bf m}\cdot{\bf H}_d
        = V\half \mu_0 M_{\rm s}^2 \hat{\bf m}\cdot{\mathcal N}\cdot\hat{\bf m},
    \label{eqn:E_shape}
\end{equation}
where ${\bf H}_d = M_{\rm s}
{\mathcal N}\cdot\hat{\bf m}$ and ${\mathcal N}$ are the demagnetization field and demagnetization
tensor, respectively.
Referring to Fig.~\ref{fig:geometry}, the total energy is 
\begin{eqnarray}
    E &=& \half \mu_0 M_{\rm s}^2 V
[L \cos^2\theta  + M \sin^2\theta\sin^2\phi \nn
        & & + N \sin^2\theta\cos^2\phi]- \mu_0 M_{\rm s} VH\cos\theta \nn
       & & - {2VK_u\over d}\sin^2\theta\cos^2\phi
\end{eqnarray}
where $L, M, N$ are the demagnetization factors for the $\hat{z}, \hat{y},
\hat{x}$ directions. These terms can be combined to give
\begin{eqnarray}
    {2 E\over \mu_0 M_{\rm s}^2V}
       &=& h_L\cos^2\theta + h_M\sin^2\theta\sin^2\phi
\nn&& + h_N\sin^2\theta\cos^2\phi
       - 2h\cos\theta,
\end{eqnarray}
where $h=H/M_s$, and
\begin{equation}
    h_L = L-{H_k\over M_{\rm s}},~~~
    h_M = M, ~~~
    h_N = N-{4K_u \over \mu_0 M^2_{\rm s} d}.
    \label{eqn:dimlessfields}
\end{equation}
If we model the thin free layer as a very flat
ellipsoid with semi-axis $a\ge b\gg c$,  Eqs.~(2.23-25) in
Ref.~\onlinecite{Osborn:1945} give
\begin{eqnarray}
    L &=& {c\over a}(1-e^2)^{1/2}{K-E\over e^2} \\
    M &=& {c\over a}{E-(1-e^2)K\over e^2(1-e^2)^{1/2}} \\
    N &=& 1 - {cE\over a(1-e^2)^{1/2}},
\end{eqnarray}
where $K$ and $E$ are complete elliptic integrals with argument
$e=\sqrt{1-b^2/a^2}$.  For the nominal geometry of
Ref.~\onlinecite{Kiselev:2003}, we have $2a=130\,{\rm nm}, 2b=70\,{\rm
nm}, 2c=3\,{\rm nm}$, so $L\approx 0.017, M\approx 0.035, N\approx
0.948$.

\end{document}